\documentstyle[sprocl,epsfig]{article}

\newcommand{\lsim}{\mathrel{\rlap{\lower4pt\hbox{\hskip1pt$\sim$}}
    \raise1pt\hbox{$<$}}}         
\newcommand{\gsim}{\mathrel{\rlap{\lower4pt\hbox{\hskip1pt$\sim$}}
    \raise1pt\hbox{$>$}}}         
\newcommand{\bq}{\begin{eqnarray}}
\newcommand{\eq}{\end{eqnarray}}
\def\VEC#1{\hbox{\boldmath$#1$\unboldmath}}
\def\Vec#1{{\bf #1}}

\def\one{1 \!\! 1}
\def\d{{\mathrm d}}
\def\e{{\mathrm e}}
\def\I{{\mathrm i}}

\begin{document}

\title{UNDERSTANDING TRANSVERSITY: \\ PRESENT AND 
FUTURE\footnote{Plenary talk at SPIN 2004 (Trieste, Oct. 2004), 
to be published in the Proceedings.}}

\author{VINCENZO BARONE}

\address{Di.S.T.A., Universit{\`a} 
del Piemonte Orientale ``A.~Avogadro'', \\
and INFN, Gruppo Collegato di Alessandria, \\
Via Bellini 25/G, 15100 Alessandria, Italy}

\maketitle

\abstracts{I review the present state of knowledge concerning 
transversity distributions and 
related observables. In particular, I discuss the phenomenology 
of transverse asymmetries in $e p^{\uparrow}$, 
$p p^{\uparrow}$, $p^{\uparrow} p^{\uparrow}$ and  
$\bar p^{\uparrow} p^{\uparrow}$
scattering, and the perspectives of ongoing and future research.}

\section{General properties of transversity}
\label{intro}

The transverse polarization, or {\em transversity}, distribution 
of quarks $h_1(x)$ -- also called $\Delta_T q(x)$ -- 
has been the 
subject of an intense theoretical work in the last decade 
(for reviews, see Refs.~1 and 2), and the corresponding observables 
are now actively investigated in many experiments. 

Let us start by recalling the partonic definition of $h_1(x)$.
Given a transversely polarized hadron, if we denote by $q_{\uparrow} 
(q_{\downarrow})$ 
the number
density of quarks with polarization parallel (antiparallel) to that
of the hadron, the transversity distribution is the difference 
$h_1(x) = q_\uparrow(x) - q_\downarrow(x)$. 
In field-theoretical terms, 
$h_1(x)$ is given by ($P$ and $S$ are the momentum and the spin of the 
hadron, respectively)
\begin{equation}
 h_1(x) = \int \frac{\d \xi^-}{4 \pi} 
\, \e^{\I x P^+ \xi^-}  \!
\left. \langle P, S \vert \overline{\psi}(0) \gamma^+ \gamma_{\perp} 
\gamma_5 \psi(\xi) 
\vert P, S \rangle \right \vert_{\xi^+ = \Vec \xi_{\perp} = 0}\,,  
\label{eq2}
\end{equation} 
and is a {\em leading-twist} quantity, like the number density 
$f_1(x)$ (also called $q(x)$) and the helicity distribution 
$g_1(x)$ (more often, and less ambiguously, called $\Delta q(x)$).  
A Wilson line $W(0, \xi)$ should be inserted between the 
quark fields in (\ref{eq2}), in order to ensure gauge invariance. In the 
light-cone gauge, $W$ reduces to unity
and can be omitted (this is no more true for $k_T$-dependent 
distributions, see below). The tensor charge $\delta q$ is defined by 
\begin{equation}
 \langle P, S \vert \overline{\psi}(0) 
\I \sigma^{\mu \nu} \gamma_5 \psi(0) \vert P, S \rangle 
= 2 \, \delta q \, S^{[\mu} P^{\nu ]}\,, 
\label{eq3}
\end{equation} 
and corresponds to the first moment of $h_1 - \bar h_1$: 
$\delta q = \int \d x \, (h_1^q - \overline{h}_1^q )$.    

An important peculiarity of $h_1$ 
is that it has no gluonic
counterpart (in spin-$1/2$ hadrons). Therefore, 
it does not mix 
with gluons, and behaves as a non-singlet distribution.
At low $x$, it turns out to rise slower than
$g_1$ by QCD evolution.\cite{Barone:1997fh} 
An angular momentum sum rule for transversity,  
\begin{equation}
\frac{1}{2} = \frac{1}{2} \, \sum_{a = q, \bar q} \int \d x \, 
h_1^a(x) + \sum_{a = q, \bar q, g} \langle L_T 
\rangle^a \,, 
\label{eq6}
\end{equation}
has been recently proven in the framework 
of the quark-parton model.\cite{Bakker:2004ib} 
Since transversity decreases with increasing $Q^2$, 
the orbital angular momentum $\langle L_T \rangle$  
must increase
(assuming an initial zero value). Of course, 
it would be 
very interesting to study this sum rule in 
perturbative QCD.

The transversity distributions have been computed in 
a variety of models (for a review, see Ref.~1). 
Generally, one finds $h_1 \approx g_1$ at the model scale,  
i.e. for $Q^2 \lsim 0.5$ GeV$^2$ (the difference between the 
two quantities comes from the lower components of the 
quark wavefunctions). 
Tensor charges have been also evaluated in lattice QCD 
and by QCD sum rules. 
A summary of all estimates is:
$\delta u \sim 0.7-1.0, \, 
\delta d \sim - (0.1 - 0.4)$  at 
$Q^2 = 10 \; {\rm GeV}^2$.

Examining the operator structure in (\ref{eq2}) one sees that 
$h_1(x)$ is \emph{chirally-odd}. Now, fully inclusive
DIS proceeds via the so-called handbag diagram, which cannot flip the
chirality of quarks. Thus, transversity
distributions are {\em not} observable in inclusive DIS. 
In order to measure $h_1$, the
chirality must be flipped twice, so one needs either two hadrons in the initial
state (hadron--hadron collisions), or one hadron in the initial state and 
one - at least - 
in the final state (semi-inclusive deep inelastic 
scattering, SIDIS).

\section{$k_T$-dependent distributions related to transversity} 
\label{kT}

If we ignore (or integrate over) the transverse 
momenta of quarks,  
$f_1(x)$, $g_1(x)$ and $h_1(x)$ completely describe the
internal dynamics of hadrons. 
Taking the transverse motion of quarks into account, 
the number of distribution functions increases. At
leading twist there are eight
$k_T$-dependent distributions\cite{Mulders:1996dh,Kotzinian:1994dv}, 
three of which, upon integration over $\Vec{k}_T^2$, yield $f_1(x)$, 
$g_1(x)$ and $h_1(x)$. 
The remaining five distributions
are new and disappear when the hadronic tensor is integrated over
$\Vec{k}_T$. They are related to various 
correlations between $\Vec k_T$, $\Vec S_T$ and 
$\Vec S_{qT}$ (the quark spin). 
The spin asymmetry of transversely polarized quarks inside a 
transversely polarized proton is given by 
\bq 
{\mathcal P}_{q^{\uparrow}/p^{\uparrow}}(x, \Vec k_T) - 
{\mathcal P}_{q^{\downarrow}/p^{\uparrow}}(x, \Vec k_T)
& =&  (\Vec S_T \cdot \Vec S_{qT}) \,  h_1 (x, \Vec k_T^2)
\nonumber \\
& & \hspace{-3.5cm} 
-  \frac{1}{M^2} \left [ (\Vec k_T \cdot \Vec S_{qT}) (\Vec k_T 
\cdot \Vec S_T) + \frac{1}{2} \, \Vec k_T^2 (\Vec S_T \cdot \Vec S_{qT})
\right ]  h_{1T}^{\perp} (x, \Vec k_T^2)\,, 
\eq
and contains not only the unintegrated transversity 
distribution $h_1(x, \Vec k_T^2)$, but also another
distribution function, called $h_{1T}^{\perp} (x, \Vec k_T^2)$. 
Both $h_1$ and $h_{1T}^{\perp}$ contribute to single-spin asymmetries 
 in SIDIS (via Collins effect),  
but with different angular distributions, $\sin (\phi_h + \phi_S)$ 
and $\sin (3 \, \phi_h - \phi_S)$ respectively (see below). 
Consider now unpolarized quarks inside a transversely polarized 
proton. They may have an azimuthal asymmetry of the form 
\begin{equation}
{\mathcal P}_{q/p^{\uparrow}}(x, \Vec k_T) - 
{\mathcal P}_{q/p^{\uparrow}}(x, - \Vec k_T)
 = \frac{(\Vec k_T \times \hat{\Vec P}) \cdot \Vec S_T}{M} 
\, f_{1T}^{\perp} (x, \Vec k_T^2)\,, 
\end{equation}
where $f_{1T}^{\perp}$ is the {\em Sivers distribution 
function}\cite{Sivers}.  
Specularly, transversely polarized quarks inside an unpolarized 
proton admit a possible spin asymmetry of the form
\begin{equation}
{\mathcal P}_{q^{\uparrow}/p}(x, \Vec k_T) - 
{\mathcal P}_{q^{\downarrow}/p}(x, \Vec k_T)
 = \frac{(\Vec k_T \times \hat{\Vec P}) \cdot \Vec S_{qT}}{M} 
\, h_1^{\perp} (x, \Vec k_T^2)\,, 
\end{equation} 
where $h_1^{\perp}$ is the so-called {\em Boer--Mulders distribution 
function}\cite{Boer:1998nt}. 
The two distributions $f_{1T}^{\perp}$ 
and $h_1^{\perp}$ are associated with the {\em T-odd} correlations  
$(\hat{\Vec P} \times \Vec k_T) \cdot \Vec S_{T}$ and 
$(\hat{\Vec P} \times \Vec k_T) \cdot \Vec S_{qT}$. To see 
the implications of time-reversal invariance, let us write
the operator definition of the Sivers function: 
\begin{eqnarray}
f_{1T}^{\perp} (x, \Vec k_T^2)  &\sim& 
\int \d \xi^- 
\int \d \Vec \xi_T \, 
\e^{\I x P^+ \xi^- - \I k_T \cdot  \xi_T}
\nonumber \\
& & \;\;
\times \,  \langle P, S_T \vert \overline{\psi}(\xi) \gamma^+ 
W(0, \xi) 
\psi(0) 
\vert P, S_T \rangle 
\label{sivfun1}
\end{eqnarray}
If we na\"{\i}vely set the Wilson link $W$ to $\one$,  
the matrix element in (\ref{sivfun1}) 
 changes sign 
under time reversal $T$, hence 
the Sivers function 
must be zero.\cite{Collins:1993kk}  
On the other
hand, a direct
calculation\cite{Brodsky:2002cx} in a quark-spectator model 
shows that $f_{1T}^{\perp}$ is non vanishing: gluon exchange 
between the struck quark and the target remnant generates 
a non-zero Sivers asymmetry
(the presence of a quark transverse momentum smaller than
$Q$ ensures that this asymmetry is proportional to $M/k_T$, rather than to
$M/Q$, and therefore is a leading-twist observable).
The puzzle is solved by 
carefully considering the Wilson line in 
(\ref{sivfun1}).\cite{Collins:2002kn} For the case at hand 
(SIDIS), $W(0, \xi)$ includes a link at $\infty^-$
which does not reduce to $\one$ in the 
light-cone gauge.\cite{Belitsky} 
Time reversal changes 
 a future-pointing Wilson line 
 into a past-pointing Wilson line and therefore invariance
under $T$, rather than constraining $f_{1T}^{\perp}$ to zero,
gives a relation between processes that probe Wilson lines 
pointing in opposite time directions. 
In particular, since in SIDIS the Sivers asymmetry arises from the interaction 
between the spectator and the outgoing quark, whereas in Drell-Yan 
processes it is due to the interaction between the spectator 
and an incoming quark, one gets
$f_{1T}^{\perp} (x, \Vec k_T^2)_{\rm SIDIS} = - 
f_{1T}^{\perp} (x, \Vec k_T^2)_{\rm DY}$.  
This is an example of the ``time-reversal modified
universality'' of distribution functions in SIDIS, Drell-Yan 
production and $e^+ e^-$ annihilation studied by Collins and 
Metz.\cite{Collins:2004nx} More complicated Wilson link 
structures in various hard processes have been investigated by 
Bomhof, Mulders and Pijlman.\cite{Bomhof:2004aw} The issue 
is not completely settled and more theoretical work seems 
to be needed in order to fully clarify the universality 
properties of $k_T$-dependent distributions.   
Finally, it is known\cite{Teryaev} that at twist 3, effective $T$-odd 
distributions emerge from gluonic poles. The precise connection 
between $k_T$-dependent and twist-3 distributions is 
another problem that deserves further study.

\section{Probing transversity}

\subsection{Semi-inclusive deep inelastic scattering}

Let us start with the
single-spin process $e \, p^{\uparrow}\, 
\to\, e'\, \pi\, X$, for which some data are already
available. 
In order to have a non vanishing 
asymmetry, one must consider the transverse motion 
of quarks. The non-collinear factorization 
theorem has been recently proven by Ji, Ma and Yuan\cite{Ji:2004xq} for
$P_{hT} \ll Q$.
A single-spin transverse asymmetry is due either to: {\it i)}  
a spin asymmetry of transversely 
polarized quarks fragmenting into the unpolarized hadron, the 
so-called {\em Collins effect} involving 
\begin{equation}
 {\mathcal N}_{h/q^{\uparrow}}(z, \Vec P_{hT}) - 
{\mathcal N}_{h/q^{\downarrow}}(z, \Vec P_{hT}) = 
\frac{( \hat{\VEC \kappa}_T \times 
\Vec P_{hT}) \cdot \Vec S_{qT}}{zM_h} 
\, H_1^{\perp} (z, \Vec P_{hT}^2) \,, 
\label{funz_collins}
\end{equation}
a $T$-odd function 
not forbidden by time reversal invariance (due to final-state 
interactions);  
or to {\it ii)}
an azimuthal asymmetry  
of unpolarized quarks inside the transversely polarized proton,  
the so-called {\em Sivers effect}, involving $f_{1T}^{\perp}$.  
The differential cross section for $e \, p^{\uparrow}\, 
\to\, e'\, \pi\, X$ is
\bq
\d \sigma 
&\sim&   
A(y) \, {\mathcal I} \left [ \frac{\VEC \kappa_T \cdot 
\hat{\Vec P}_{h T}}{M_h} \, h_{1} \, H_{1}^{\perp} 
\right ]\, \sin (\phi_h + \phi_S) 
\nonumber \\
& & 
 + \, 
B(y) \, {\mathcal I} \left [ \frac{\Vec k_T \cdot 
\hat{\Vec P}_{h T}}{M_h} \, f_{1T}^{\perp} \, D_{1} 
\right ]\, \sin (\phi_h - \phi_S) 
\nonumber \\
& &  
+ \, 
C(y) \, {\mathcal I} \left [ 
\lambda (\Vec k_T, \VEC \kappa_T, \hat{\Vec P}_{h T}) \, 
 h_{1T}^{\perp } \, H_{1}^{\perp } \right ] \, 
\sin (3 \phi_h - \phi_S) \,, 
\eq
where ${\mathcal I} [\ldots]$ is a
 convolution integral 
over $\Vec k_T$ and $\VEC \kappa_T$. 
As one can see, there is a variety of angular distributions
which combine in different ways the two physical angles 
$\phi_h$ and $\phi_S$. In particular, the Collins effect 
is associated with $\sin (\phi_h + \phi_S)$, and also 
with $\sin (3 \phi_h - \phi_S)$ if the transverse motion 
of quarks inside the target is not neglected, whereas 
the Sivers effect is associated with a $\sin (\phi_h - \phi_S)$
distribution. One can disentangle these angular 
distributions by taking the azimuthal moments of the asymmetries.  
For instance, the Collins moment is 
\begin{equation}
\langle \sin (\phi_h + \phi_S) \rangle \equiv 
\frac{\int \d \phi_h \, \d \phi_S \, \sin (\phi_h + \phi_S) 
\, [\d \sigma (\phi_h, \phi_S) - 
\d \sigma (\phi_h, \phi_S + \pi)]}{
\int \d \phi_h \, \d \phi_S \,
[\d \sigma (\phi_h, \phi_S) +
\d \sigma (\phi_h, \phi_S + \pi)]}. 
\nonumber 
\end{equation}
Recently, the HERMES Collaboration\cite{Airapetian:2004tw} reported 
the first measurement
of the Collins 
moment $\langle \sin (\phi_h + \phi_S) \rangle$ and of  
the Sivers moment $\langle \sin (\phi_h - \phi_S) \rangle$,
in the region $0.02 < x < 0.4$, $0.2 < z < 0.7$, 
at $\langle Q^2 \rangle = 2.4$ 
GeV$^2$. The Collins asymmetry  
$A_T^{\pi^+}$ is found to be positive, whereas $A_T^{\pi^-}$ is negative.
This is consistent with the fact that   
$h_1^u > 0$ and $h_1^d < 0$. However,   $A_T^{\pi^-}$ 
is negative and its absolute value  
$\vert A_T^{\pi^-} \vert$ is larger than  
$\vert A_T^{\pi^+} \vert$, 
whereas one expects from models $\vert h_1^d \vert \ll \vert h_1^u \vert$. 
Recalling that $A_T^{\pi^+}$ and $A_T^{\pi^-}$ involve the following 
combinations of distribution and fragmentation functions 
(`fav' = favored, `unf' = unfavored)
\begin{equation}
A_T^{\pi^+}: \;\;\; 4 \, h_1^u \, H_1^{\perp {\rm fav}} + 
 h_1^d \, H_1^{\perp {\rm unf}}\,, \hspace{0.5cm} 
A_T^{\pi^-}: \;\;\; h_1^d \, H_1^{\perp {\rm fav}} + 
 4 \, h_1^u \, H_1^{\perp {\rm unf}}  
\end{equation}
one sees that the $\pi^-$ data 
seem to require large unfavored Collins functions, 
with $H_1^{\perp {\rm unf}} \approx - H_1^{\perp {\rm fav}}$. 
It would be very useful to get some independent 
information on $H_1^{\perp}$ from other processes: in this respect,  
the forthcoming extraction of $H_1^{\perp}$ from 
 $e^+ e^-$ annihilation data in the Belle experiment at KEK 
will be extremely important.\cite{Seidl} 
There are also preliminary HERMES results on the $\pi^0$ 
asymmetry, showing a largely negative $A_T^{\pi^0}$, 
similar to $A_T^{\pi^-}$. This is quite a controversial finding, 
as it conflicts with expectations based on isospin invariance. 
The Collins asymmetry has also been measured 
by the COMPASS Collaboration with a deuteron target.\cite{Pagano} 
In the  $x \lsim 0.1$ region, it is found to be compatible with 
zero for both $\pi^+$ and $\pi^-$, as expected 
quite generally at small $x$.  
Concerning Sivers asymmetries, HERMES find $A_T^{\pi^+} > 0$: 
this is the first evidence of a non vanishing Sivers function 
$f_{1T}^{\perp}$, although -- due to the smallness 
of $A_T^{\pi^+}$ and the present uncertainties -- more precise data 
are needed to draw a definite conclusion.

Another access to transversity in the context of SIDIS is offered 
by the double-spin process $e \, p^{\uparrow}\, \to\, e'\, \Lambda^{\uparrow}
\, X$ (transversely polarized $\Lambda$ production), which probes 
the fragmentation analog of $h_1$, i.e.
$H_1 (z) = {\mathcal N}_{h^{\uparrow}/q^{\uparrow}}(z) - 
{\mathcal N}_{h^{\uparrow}/q^{\downarrow}}(z)$. 
Unfortunately, it is hard to predict the $\Lambda$  
polarization, 
because $H_{1 \Lambda}$ is 
unknown (see, however, some attempts  
in Ref.~20). An analysis of data on transversely  
polarized $\Lambda$ production is currently being 
performed by the COMPASS Collaboration.\cite{Bressan}  

A third promising process to detect transversity  
is two-pion production in $e p^{\uparrow}$ scattering. 
In this case, after integrating the cross sections 
over $\Vec P_{hT}$, one finds that the single-spin asymmetry 
depends on an interference fragmentation function $I(z, M_h^2)$, 
arising from the interference between different partial waves 
of the two-pion system.\cite{twopion} The extraction 
of this function is under way.\cite{Joosten}

\subsection{Pion hadroproduction} 

Collins and Sivers effects manifest themselves 
also in pion hadroproduction  
with a transversely polarized target. 
A non vanishing 
asymmetry is generated either by quark transverse momenta 
or by higher-twist effects. 
The non-collinear factorization formula 
is, in this case, only conjectured. Assuming its validity, 
the Collins asymmetry reads
\bq
\d \sigma^{\uparrow} - \d \sigma^{\downarrow} &\sim& 
\sum_{abc}   [h_1(x_a, \Vec k_T^2) + (\Vec k_T^2/M^2) \, 
h_{1T}^{\perp} (x_a, \Vec k_T^2)] \, 
 \otimes \, f_1(x_b, {\Vec {k}'_T}^2) 
\nonumber \\
& & \hspace{0.5cm} \otimes \, 
  \Delta_{TT} \hat\sigma(a^{\uparrow} b \to c^{\uparrow} d) 
 \, \otimes \,  H_1^{\perp}(z, \Vec\kappa_T^2) \,,
\eq
whereas the Sivers asymmetry is
\begin{equation}
\d \sigma^{\uparrow} - \d \sigma^{\downarrow} \sim 
  f_{1T}^{\perp} (x_a, \Vec k_T^2) 
\, \otimes \, f_1(x_b, {\Vec {k}'_T}^2) \, \otimes \,  
  \d \hat\sigma(a b \to c d) \, 
 \otimes \,  D_1 (z, \VEC \kappa_T^2) \,. 
\end{equation}
An extensive and detailed treatment of single-spin asymmetries 
in the framework of non-collinear factorization has 
been presented in Refs.~24 and 25 (for another approach 
leading to similar conclusions, see Ref.~26). The main finding is that 
the Collins asymmetry alone is unable to reproduce the 
E704\cite{Adams:1991} and STAR\cite{Adams:2003fx}
data: the Collins effect turns out to be 
suppressed due to kinematic phases occurring in non-collinear 
partonic subprocesses (this does not imply anything about the 
magnitude of $H_1^{\perp}$). On the contrary, the Sivers 
mechanism is not affected by a similar suppression.
A major shortcoming of pion hadroproduction is that it 
depends on one physical angle only, so that all
asymmetry mechanisms are entangled. A possible way 
to avoid this problem is to study a less inclusive 
process, such as pion + jet production, 
as advocated by Teryaev (private communication).  
 
In a recent paper, Bourrely and Soffer\cite{Bourrely:2003bw} 
argued that, since 
collinear pQCD correctly reproduces the large-$\sqrt{s}$ 
STAR unpolarized cross sections but fails 
to describe the small-$\sqrt{s}$ E704 data, the 
single-spin asymmetries measured by these two experiments are 
actually different phenomena, and in particular the E704 
asymmetry ``cannot be attributed to pQCD''. Two comments are 
in order: first of all, higher-twist effects might be important,  
since $\langle P_{hT} \rangle$ is not so large (typically, around 
1-2 GeV). Second, as shown by D'Alesio and 
Murgia\cite{D'Alesio:2004up}, quark transverse momenta 
considerably improve the agreement of the pQCD   
calculations with the small-$\sqrt{s}$ unpolarized cross sections.

\subsection{Drell-Yan processes} 

\begin{figure}[t]
\begin{center}
\hspace{-6cm}
\parbox{6cm}{
\scalebox{0.38}{
\includegraphics*[70,450][530,750]{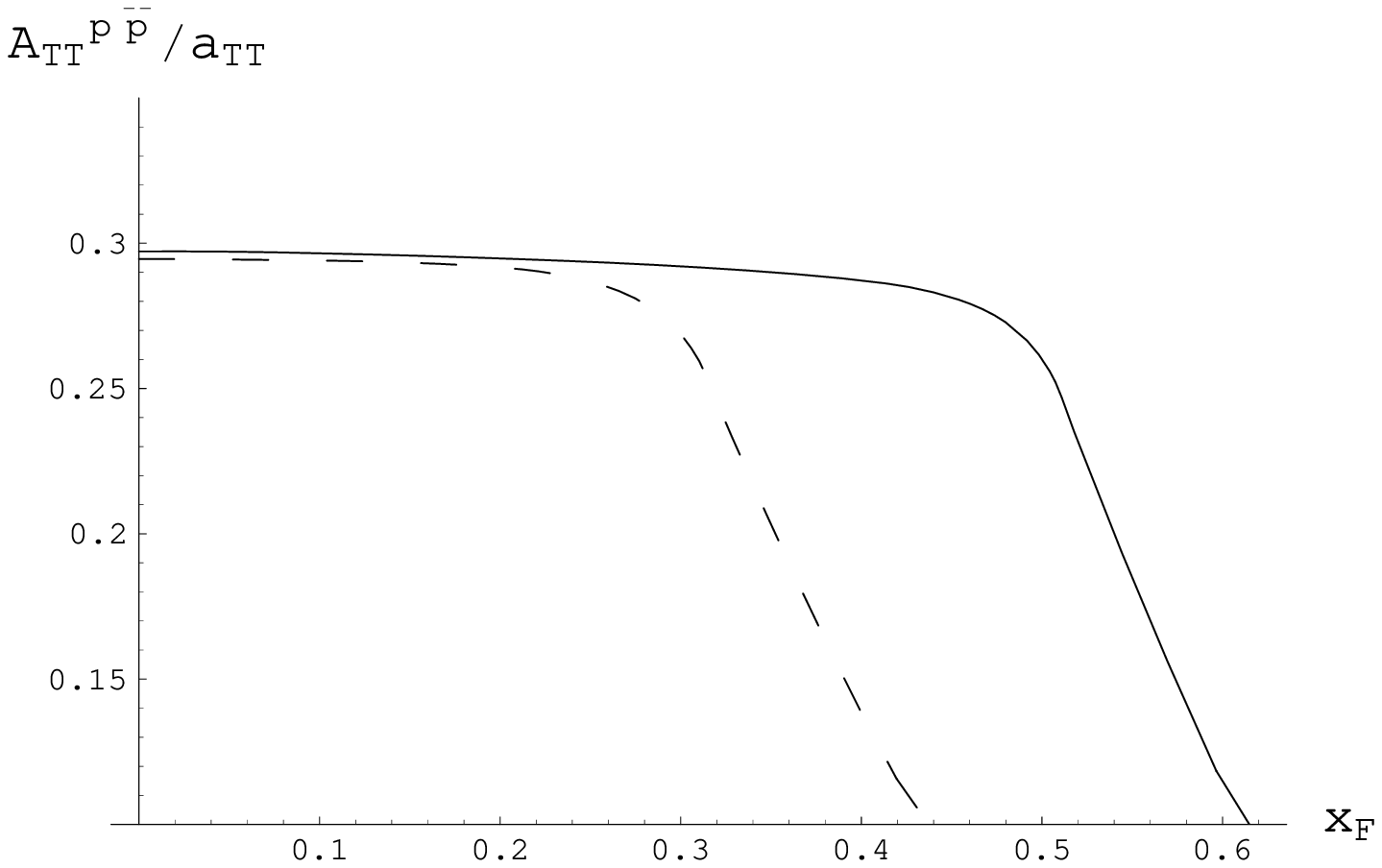}
\hspace{-1cm} 
\includegraphics*[70,450][530,750]{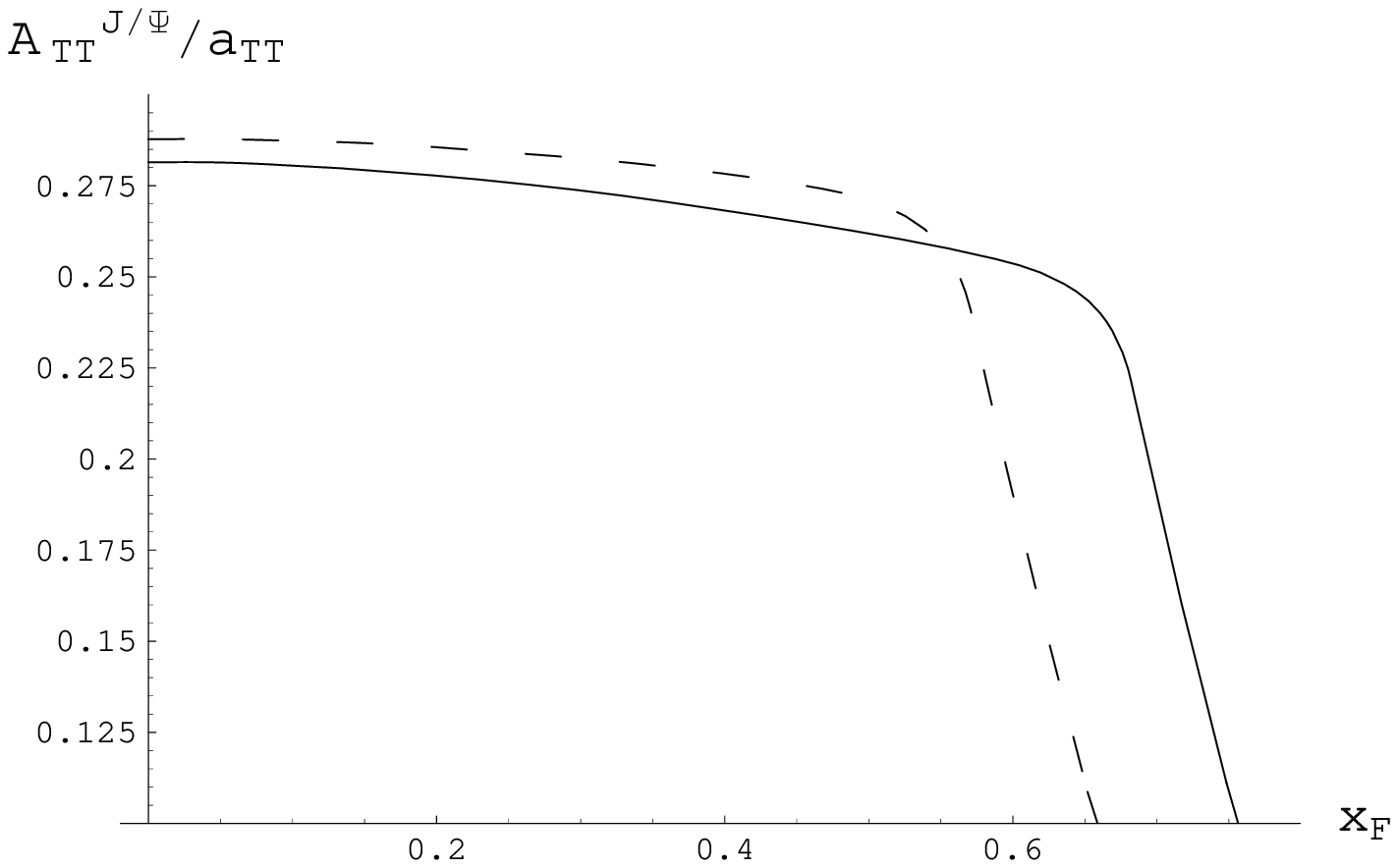}
}}
\end{center}
\caption{Transverse double spin asymmetry in $\bar p^{\uparrow} 
p^{\uparrow}$ Drell-Yan 
production at $M = 4$ GeV (left) and for $J/\psi$ production (right), as 
a function of $x_F = x_1 - x_2$. 
Solid (dashed) lines correspond to $s = 45 \, (30)$ GeV$^2$.}  
\label{fig_att}
\end{figure}

Drell-Yan production in $p^{\uparrow} p^{\uparrow}$ collisions is 
the cleanest process that probes transversity. 
The double-spin asymmetry $A_{TT}^{\rm DY}$, in fact, contains 
only combinations of the transversity distributions.  
At leading order, for instance, one has 
\begin{equation}
A_{TT}^{DY} (pp) \sim 
\frac{\sum_q e_q^2 h_1^q(x_1, M^2) 
 \bar h_1^q(x_2, M^2) + [1 \leftrightarrow 2]}{\sum_q e_q^2 f_1^q(x_1, M^2) 
 \bar f_1^q(x_2, M^2) + [1 \leftrightarrow 2]}\,. 
\label{att_pp}
\end{equation} 
It turns out, however,  
that at the energies of 
RHIC (where this process will be studied\cite{Saito}) 
this asymmetry is rather small\cite{Barone:1997mj,Martin}
 (about $1-2 \, \%$; similar values are found for 
transverse double-spin asymmetries in prompt photon 
production\cite{Mukherjee:2003pf}).   
The reason is twofold: 1) $A_{TT}^{DY} (pp)$ contains 
antiquark transversity distributions, which are 
small; 2) RHIC kinematics ($\sqrt{s} = 200$ GeV, $M < 10$ GeV, 
$x_1 x_2 = M^2/s \lsim 3 \times 10^{-3}$)
probes the low-$x$ region, where $h_1$ rises slowly. 
The problem 
could be circumvented by considering $\bar p^{\uparrow} p^{\uparrow}$ 
scattering at more moderate energies. In this case a much 
larger asymmetry is 
expected\cite{Barone:1997mj,Anselmino:2004ki,Efremov:2004qs} 
since $A_{TT}^{DY} (\bar p p)$ 
contains products of valence distributions 
at medium $x$. The PAX Collaboration 
has proposed to study $\bar p^{\uparrow} p^{\uparrow}$-initiated 
Drell-Yan production at the High-Energy Storage Ring of GSI, 
in the kinematic region   
$30 \, {\rm GeV}^2 \lsim s \lsim 45 \, {\rm GeV}^2, 
\, M \gsim 2 \, {\rm GeV}, \, x_1 x_2 \gsim 0.1$.\cite{Rathmann}  
Leading-order predictions for the $\bar p p$ asymmetry in this 
regime are shown in Fig.~\ref{fig_att} (left). 
$A_{TT}^{DY} (\bar p p)$ is as large as 
0.3 at $M = 4$ GeV, but counting 
rates are small and this makes the measurement arduous. Things 
become easier if one looks at the $J/\psi$ peak, where 
the production rate is larger by two orders of magnitude. 
Assuming the dominance of $q \bar q$ fusion (as suggested 
by a comparison of $pp$ and $\bar p p$ cross sections at the 
CERN SPS), the $J/\psi$ production double transverse asymmetry 
$A_{TT}^{J/\psi} (\bar pp)$ has the same structure 
as Eq.~(\ref{att_pp}), with the electric charges replaced by 
the $q \bar q - J/\psi$ couplings. 
These cancel out in the ratio, and hence
$A_{TT}^{J/\psi} (\bar pp)$, which is dominated by the $u$ sector,  
becomes
\begin{equation}
A_{TT}^{J/\psi} (\bar pp) \sim 
\frac{h_1^u(x_1, M_{J/\psi}^2)
 h_1^u(x_2, M_{J/\psi}^2)}{f_1^u(x_1, M_{J/\psi}^2) 
f_1^u(x_2, M_{J/\psi}^2)}\,. 
\end{equation} 
This asymmetry is also of the order of 0.3 (Fig.~\ref{fig_att}, right) 
and, by measuring it, 
one can directly extract the $u$ transversity distribution.

\section{Conclusions and perspectives}

Transversity is presently a very hot topic in high-energy 
spin physics. From the theoretical point of view, 
a lot of work has been done and $h_1$ 
is by now rather well known.
On the experimental
side, the era of data has at last arrived: single-spin processes 
are under study and the first results are already 
matter of phenomenological analyses. Double-spin processes 
are experimentally more difficult but theoretically cleaner, 
and their investigation is certainly worth the effort.  
While we look forward to more - and more precise - data, 
the main goal of theory is to achieve a solid   
picture of single-spin transverse asymmetries (sheding 
further light on $k_T$ and higher-twist effects), 
in view of future global studies of transversity measurements.

\section*{Acknowledgments}

I am grateful to the organizers of SPIN04  
for their invitation to this beautiful Symposium, and to 
M.~Anselmino, 
U.~D'Alesio, A.~Efremov, L.~Gamberg, 
A.~Kotzinian, A.~Prokudin, 
P.~Ratcliffe and O.~Teryaev for useful 
discussions.


\end{document}